\documentclass[11pt]{article}
\usepackage{times}
\usepackage{geometry}
\usepackage[utf8]{inputenc}
\usepackage{enumitem,amssymb}
\usepackage{graphicx}
\usepackage{fancyhdr}
\usepackage{aas_macros}
\usepackage{mdframed} 
\usepackage{hyperref,url}

\usepackage[authoryear]{natbib}
\setcitestyle{authoryear,open={(},close={)}}

\mdfdefinestyle{theoremstyle}{
innertopmargin=\topskip,}
\mdtheorem[style=theoremstyle]{lrptextbox}{}

\newcommand{\tcm}{$21\,\textrm{cm}\,\,$}  

\begin{document}
\title{Low-redshift 21cm Cosmology in Canada}
\author{A. Liu, S. Foreman, H. Padmanabhan, H. C. Chiang, S. Siegel, D. Wulf, J. Sievers, M. Dobbs, K. Vanderlinde}
\date{Submitted to LRP2020 panel on 30th September 2019}

\maketitle

%


\section{Introduction}

While traditional cosmological probes such as the Cosmic Microwave Background (CMB) and galaxy surveys have provided highly precise constraints on $\Lambda$CDM cosmology, new insights into beyond-$\Lambda$CDM physics may require new probes. Line-intensity mapping of the \tcm line has the potential to be just such a probe, with its promise to measure larger portions of our observable Universe than are accessible via traditional means, at a fraction of the cost. In this white paper, we argue that Canada has been a leader in low-redshift \tcm cosmology, having used existing instruments to produce world-leading constraints and having built up a set of first-generation instruments that are now taking data. Now is thus the time to capitalize on this foundation to ensure continued leadership as low-redshift \tcm cosmology moves from upper limits to science.

In this white paper, we concentrate on the application of \tcm cosmology at $z < 6$, i.e., after the Epoch of Reionization. (Please see companion white paper E011 ``High-redshift \tcm  Cosmology in Canada" for a discussion of the opportunities at $ z>6$). At $z<6$, large scale structure is traced by dense self-shielded clumps of neutral hydrogen within galaxies, and therefore can be observed by performing surveys of the \tcm line over large areas of the sky. This can be done relatively inexpensively using radio telescopes of moderate resolution because it is unnecessary to resolve individual galaxies. Instead, one can simply search for the total integrated emission over rather coarse pixels in the sky. Doing so has two advantages over traditional large-scale structure surveys, where one identifies individual bright objects (e.g. galaxies) and uses them as tracers of the matter distribution. The first advantage is that by measuring the total integrated emission, we are making use of every photon along the luminosity function, including emission from objects that would be too faint to see in a traditional survey. Second, individual astronomical objects tend to have characteristic length scales that are much smaller than cosmological scales. If one does not need to pick out the individual objects, one can design instruments that directly probe the scales of interest, resulting in a more efficient survey.

\begin{figure*}[h!]
\centering
\includegraphics[width=0.75\textwidth,trim={0cm 0cm 0cm 0cm},clip]{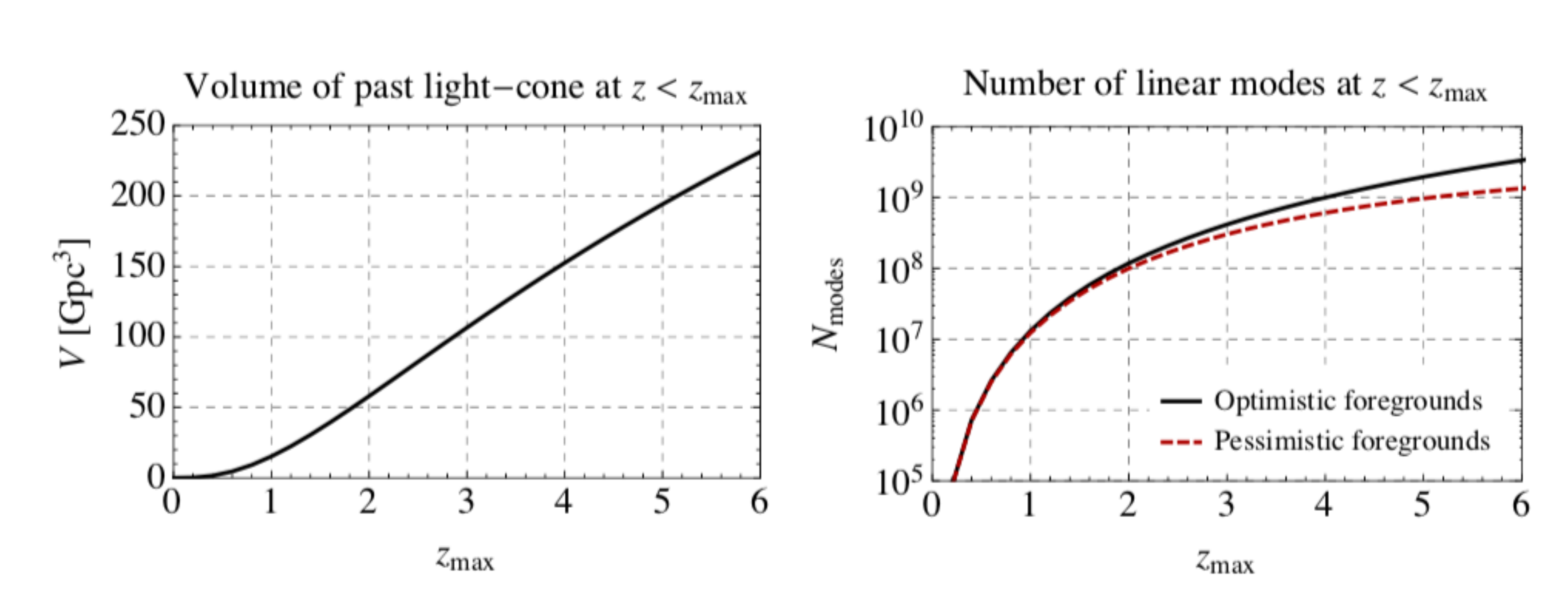}
\caption{Left: Total comoving volume of our Universe from $z=0$ to various values of $z_\textrm{max}$. Right: The corresponding number of easy-to-model linear modes that are in principle measurable within the volume. The presence of foreground contaminants necessitate cutting out certain Fourier modes, and so fewer modes are available for cosmology if one makes more pessimistic assumptions about one's ability to remove foregrounds. From \citet{Ansari:2018}}
\label{fig:modes}
\end{figure*}

Along with the aforementioned advantages, the \tcm line also provides access to redshifts that are difficult for other cosmological probes to reach. This enables \tcm instruments to probe unprecedentedly large volumes of our Universe, as illustrated in Figure \ref{fig:modes}. The corresponding number of modes accessible in such large volumes is orders of magnitude larger than with current cosmological probes, enabling exquisite constraints on standard $\Lambda$CDM parameters. Alternatively, one could \emph{test} rather than \emph{assume} $\Lambda$CDM with a \tcm dataset, for example by using a long lever arm in redshift to test early dark energy models.

The use of the \tcm line to probe large-scale structure is therefore one of considerable promise. The field is at an inflection point, with zeroth-generation instruments having successfully traced large-scale structure via cross correlations, and first-generation instruments now taking data in an attempt to measure an auto-correlation signal. Canadian leadership has been crucial in these efforts, and we call for a sustained investment in the next decade to further expand our leadership role just as \tcm cosmology makes the transition from a probe with considerable theoretical promise to one that is a workhorse of cosmology.

\section{The Current Landscape for $21\,\textrm{cm}$ Cosmology at $z<6$}

The last decade has seen considerable progress in the field of \tcm cosmology at $z<6$ , both in terms of instrumentation and observations:
\subsection{Observational Results}
\label{sec:ObsResults}
Perhaps unsurprisingly, the earliest constraints from low-redshift \tcm cosmology have come from existing instruments. In 2010, \citet{Chang2010} used the Green Bank Telescope (GBT) to make a $\sim 4\sigma$ detection of the \tcm signal at $z \sim 0.8$ in cross-correlation with the DEEP2 optical galaxy survey. This was improved upon by \citet{Masui2013GBT} to a $\sim 7.4\sigma$ significance, again using data from GBT, but this time cross correlated with the WiggleZ survey. Recently, \citet{Anderson2018} used radio \tcm data from Parkes radio telescope and made a $5.7\sigma$ detection in cross correlation with the 2dF galaxy survey, at $z \sim 0.08$.

In addition to being achievements in their own right, these detections have also enabled the first scientific constraints from \tcm cosmology. In \citet{Switzer2013GBT}, it was pointed out that cross correlation measurements provide a lower limit on the \tcm signal (since galaxies and HI do not necessarily trace each other perfectly), while foreground contaminants add to any attempted \tcm autocorrelation measurement, thus converting them into upper limits. The result is that the true \tcm power spectrum must be sandwiched between these limits. Using this fact, \citet{Switzer2013GBT} were able to constrain the neutral hydrogen content at $z \sim 0.8$ to be $\Omega_{HI} b_{HI} = (0.62^{+0.23}_{-0.15})\times 10^{-3}$, where $\Omega_{HI}$ is the normalized hydrogen density and $b_{HI}$ is the HI bias.

The \citet{Anderson2018} results have been similarly interesting, and have already generated new scientific questions based on several puzzling conclusions from the data. Suppose one were to predict the amplitude of the cross power spectrum using the value of $\Omega_{HI} b_{HI}$ implied by the Arecibo Legacy Fast ALFA (ALFALFA) survey. Compared to such an expectation, the \citet{Anderson2018} measurement is low at the $15.3\sigma$ level. This suggests that either HI distributions are not as clustered relative to the matter distribution, or that the optical galaxies and HI do not trace each other as much as one might expect. Splitting the galaxy populations from 2dF into red and blue galaxies, \citet{Anderson2018} find the blue galaxies to be better correlated with HI.

We therefore see that with existing cross correlation measurements, \tcm cosmology has already been able to make scientifically interesting constraints on the state of neutral hydrogen in the low-redshift universe. These constraints provide key input into models of galaxy formation, for instance via studies of the circumgalactic medium.

\subsection{Progress in Instrumentation}

In parallel with the observational results obtained using existing facilities such as Parkes and GBT, the last decade has seen a considerable build up of new instruments for \tcm cosmology at $z < 6$. These efforts are partially in recognition of the fact that dedicated experiments are required for taking \tcm cosmology beyond the relatively local redshifts of $z \lesssim 1$, and to achieve the systematics control and sensitivity required for precision cosmology.

The Canadian Hydrogen Intensity Mapping Experiment (CHIME) is a prime example of such a dedicated instrument. It is a radio interferometer located at the Dominion Radio Astrophysical Observatory (DRAO), operating between $400$ and $800\,\textrm{MHz}$. This corresponds to a redshift range of $0.8 < z < 2.5$ for the \tcm line. CHIME consists of four large ($20\,\textrm{m} \times 100\,\textrm{m}$) cylindrical reflectors, each fitted with $256$ dual-polarization feeds. The cylindrical design of CHIME enables a large instantaneous field of view of $\sim 100^\circ$ in the north-south direction, and a narrower field of view of $\sim 1^\circ$ to $2^\circ$ in the east-west direction. For cosmology observations, the feeds are individually digitized and correlated with each other, resulting in a synthesized beam that splits up the instantaneous field of view into pixels with width $\sim 0.25^\circ$ to $0.5^\circ$. The telescope is operated as a drift-scan telescope with no moving parts, providing some much-needed stability for the control of systematics. As the sky drifts overhead, CHIME essentially maps out the entire northern sky.

In the southern hemisphere, the Hydrogen Intensity and Real-time Analysis eXperiment (HIRAX) has similar goals to CHIME. But instead of using cylinders, HIRAX will consist of up to $1024$ dishes that each have a diameter of $6\,\textrm{m}$. The instantaneous field of view of HIRAX ($\sim 5^\circ$ to $10^\circ$ within the frequency range of $400$-$800\,\textrm{MHz}$) is therefore smaller than that of CHIME. However, HIRAX will still be able to survey the entire southern sky as the dishes can be manually repointed, thus enabling a survey strategy where the telescope maps out one stripe of the sky per pointing, and multiple pointings over one or more seasons can be mosaiced into a 15,000 square degree map of nearly the whole southern sky.  HIRAX is slated to begin construction in early 2020 in the radio-quiet reserve in the South African Karoo desert, about 20 km from the MeerKAT core site.  HIRAX is already funded for more than 256 dishes, at which point it will be comparable to CHIME or the GBT in terms of collecting area. HIRAX is leveraging the significant development work put into building the CHIME back end.  The largest challenge in intensity mapping is systematic erros leaking foreground signals into the cosmology. HIRAX dishes will have different systematics than CHIME cylinders, so agreement between the two would provide powerful confirmation that any signals are cosmological.  HIRAX's southern location means that its cosmology results will be independent of CHIME's, and together they will survey over 93\% of the sky.

Construction of CHIME has proceeded rapidly over the last few years, and CHIME is now conducting full-time science-grade observations. Existing data are being analyzed with a careful view towards ensuring that systematic effects do not compromise CHIME's ability to hit its considerable design sensitivity. In parallel with these efforts, early analysis aims to detect the cosmological 21cm signal by cross-correlating CHIME data with quasar catalogs from the Sloan Digital Sky Survey (though the ultimate goal is for a detection in CHIME data alone). Additionally, CHIME has begun to produce detailed maps of the northern radio sky across its frequency band, and has discovered a considerable number of Fast Radio Bursts \citep{chimefrb2019a, chimefrb2019b, chimefrb2019c}.


\section{Canadian Leadership in the Current Landscape}
In the last decade, Canada has been a leader in low-redshift \tcm cosmology, having either led or played a key part in all of the developments highlighted in the previous section.

\subsection{Cross-correlation measurements from Parkes and GBT}
Canadian efforts were key in the $z \lesssim 1$ cross-correlation measurements between HI and galaxy surveys. Indeed, all the GBT constraints from Section \ref{sec:ObsResults} were led by graduate students and/or postdoctoral fellows at the Canadian Institute for Theoretical Astrophysics (CITA). Similarly, the Parkes cross-correlation constraints were made possible by senior CITA faculty who served in a supervisory role.

\subsection{CHIME and HIRAX}
As it name suggests, CHIME is a primarily Canadian effort. Along with a few other North American collaborators, CHIME is led by large research groups at the University of British Columbia, McGill University, and the University of Toronto.  The University of British Columbia was responsible for much of the telescope front end, including the broadband cloverleaf feed design and analog electronics, and also plays a prominent role in managing the computing infrastructure.  McGill and the University of Toronto collaborated to build the digital back end of the telescope, with McGill providing the FPGA-based digitizer and channelizer ("F-engine") and Toronto developing the GPU-based correlator ("X-Engine").  These three research groups now work together closely on instrument calibration and data analysis.

While HIRAX will physically be located in South Africa, Canada plays many leading roles.  The co-PI,hardware lead, and F-engine lead are at McGill, while Toronto/Dunlap is leading the development of the X-engine.  The FRB pipeline is lead by Perimeter Institute, and NRC is spearheading fiberglass dish development.  CITA and UBC will play leading roles in the data analysis.  Without Canadian support, expertise, and leadership, HIRAX would not have been possible.

\subsection{Theoretical Advances}
\label{ref:Theory}
In addition to leadership in observations and instrumentation, Canadian scientists have also been responsible for important theoretical advances in low-redshift \tcm cosmology in the last decade:
\begin{itemize}
\item With a large number of upcoming \tcm interferometers being designed as drift-scan telescopes that produce an extremely large amount of data, it has been crucial to determine how one could analyze the data in a computationally efficient yet optimal way (in the sense of giving the smallest possible error bars). An elegant solution to this problem was provided by then-CITA postdoctoral fellow, now-UBC scientist Dr. Richard Shaw who constructed the $m$-mode formalism for optimally reducing drift-scan telescope data \citep{Shaw2014,Shaw2015}. The $m$-mode formalism is used by CHIME as well as other telescopes seeking to do \tcm cosmology, such as the Owens Valley Radio Observatory Long Wavelength Array.
\item Many \tcm interferometers consist of extremely regular grids of dishes/antenna, driven by the desire to maximize sensitivity by having multiple baselines of the interferometer repeatedly sampling particular scales of cosmological interest. Traditional calibration algorithms are not particularly well suited to such \emph{redundant arrays}, and an alternate method known as \emph{redundant calibration} was pioneered by Prof. Adrian Liu at McGill University. This has been recently generalized to enable the inclusion of real-world systematics such as non-identical array elements and misplaced elements by Prof. Jonathan Sievers, also at McGill.
\item Because foreground contaminants are orders of magnitude larger than the cosmological signals of interest, exquisitely accurate foreground mitigation efforts are an integral part of any cosmological \tcm measurement. One danger in foreground subtraction is the possibility that foregrounds can be oversubtracted, leading to a loss of cosmological signal. Former CITA graduate students (Drs. Gregory Paciga and Kiyoshi Masui) and postdocs (Dr. Eric Switzer) have led the field in developing the formalism for quantifying---and correcting for---this signal loss \citep{Paciga2013, Switzer2015}.
\item One conservative way to deal with foregrounds is to simply ignore the Fourier modes where they are the strongest. While this approach has the potential to provide the most \emph{robust} constraints on the cosmological signal, the resulting constraints are not necessarily the most \emph{sensitive}, for foregrounds tend to reside in the largest-scale Fourier modes, which generally is where signal-to-noise is greatest. A possible remedy for this has recently been proposed by teams led by CITA faculty member Prof. Ue-Li Pen, who have pioneered tidal reconstruction techniques \citep{Zhu2018}. With these techniques, non-linear couplings between different Fourier modes can be used to reconstruct modes that have been discarded due to foregrounds from modes that are foreground-free.
\item Because line intensity maps are (in general) non-Gaussian, their information content extends beyond that which is provided by the power spectrum. One possible statistic for characterizing non-Gaussianities in a \tcm map is the voxel intensity probability distribution. Unfortunately, foreground contaminants make probability distributions difficult to measure. CITA postdoc Dr. Patrick Breysse has led the development of voxel intensity distribution statistics that use galaxy distribution information to effectively remove the influence of foregrounds \citep{Breysse2019}.
\item Simulations are a key part of any \tcm cosmology effort, for two reasons. First, there is still considerable uncertainty as to how the HI traces the mass distribution of our Universe (as evidenced by the \citealt{Anderson2018} results). Second, a believable measurement of the \tcm auto power spectrum will need to be backed up by simulations. CITA faculty member Prof. J. Richard Bond and former CITA senior fellow Marcelo Alvarez have led the \texttt{WebSky} simulations\footnote{\url{https://mocks.cita.utoronto.ca/index.php/Large_Scale_Structure_Mocks}}, which enable the relatively quick simulation of mock sky signals over the large fields of view covered by most \tcm experiments.
\end{itemize}

\section{Frontiers of $21\,\textrm{cm}$ Cosmology at $z<6$}
\label{sec:Frontiers}
The field of low-redshift \tcm cosmology is currently at an exciting time. Cross-correlation measurements with GBT and Parkes have provided proof-of-concept-level results. Dedicated instruments such as CHIME and HIRAX are taking data, enabling detailed empirical studies of systematic/hardware hurdles that need to be overcome in order for there to be a detection of the \tcm auto power spectrum within the next few years. At the same time, the availability of data allows end-to-end test runs of analysis pipelines leading up to a first detection.

The next decade will likely see progress on a number of frontiers in low-redshift \tcm cosmology:
\begin{itemize}
\item Foregrounds will remain a principal challenge of \tcm cosmology, but fortunately this is a problem that can be mitigated with better data. A key difficulty with modelling foregrounds in the $400$ to $800\,\textrm{MHz}$ range is that historically, there have not been very many high-quality surveys of the low-frequency sky. (The situation is particularly dire when one realizes that even amongst surveys that do exist, many of them do not quantify their errors!) This status quo will change dramatically with instruments like CHIME and HIRAX providing large-area sky maps at high spectral resolution within the frequency bands of interest.
\item With clues such as the \citet{Anderson2018} measurements, a better understanding of how HI traces matter should emerge. Observationally, better data will be available for new cross-correlation measurements. Instruments such as CHIME, HIRAX, and the Canadian Hydrogen Observatory and Radio transient Detector (CHORD; see Section \ref{sec:CHORD}) will improve the quality of \tcm data, while galaxy surveys such as the Dark Energy Spectroscopic Instrument (DESI) will provide larger galaxy samples that can be split in a large variety of ways based on their detailed properties. With all of this new data reaching higher redshifts than the existing cross-correlation measurements, an understanding of the large-scale behaviour of HI may additionally translate into constraints relevant to galaxy formation studies.
\item A measurement of our Universe's kinematic history via the Baryon Acoustic Oscillation (BAO) standard ruler remains an important goal of \tcm experiments. Provided systematic effects can be overcome, instruments such as CHIME have the sensitivity to make high-significance detection of BAO features from $z \sim 0.8$ to $2.5$. The 5-year CHIME survey that is currently in progress has the potential to constrain the distance-redshift relation, $D_{V}(z)/s$, to $1\%$ across this redshift range. This in turn would tighten constraints on models with an evolving dark energy equation of state, improving the Dark Energy Task Force (DETF) Figure of Merit by a factor of 4 over the value obtained from Planck and DETF Stage II experiments. While galaxy surveys will also advance significantly over the next few years and begin to access similar redshift ranges, \tcm measurements are important for two reasons. First, the systematics associated with radio measurements are likely to be highly complementary to those of optical galaxy surveys. Second, the \tcm line offers the unique ability to extend distance measurements to even higher redshifts (potentially up to reionization) via next-generation instruments such as CHORD.
\item Beyond BAO measurements and a first look at the HI-mass connection with cross correlations is a measurement of the broadband power spectrum, where the entire shape of the power spectrum (as a function of $k$ and $z$) is leveraged for its cosmological information. To do so, however, requires being able to overcome degeneracies between astrophysics and cosmology \citep{padmanabhan2015}. 
A data-driven, halo model framework \citep{padmanabhan2017a, padmanabhan2017b} can be used to determine the extent to which astrophysical uncertainties propagate into cosmological constraints in HI intensity mapping surveys. This uses a prescription to connect HI mass, $M_{\rm HI}$, with dark matter halo mass $M$, at different redshifts $z$:
    \begin{equation}
M_{\rm HI} (M,z) = \alpha f_{\rm H,c} M \left(\frac{M}{10^{11} h^{-1} 
M_{\odot}}\right)^{\beta_1} \exp\left[-\left(\frac{v_{\rm{c,0}}}{v_c(M,z)}\right)^3\right] 
\label{eq:HIhalomodel}
\end{equation}
with three free parameters: $\alpha$, the overall fraction of HI relevant to the cosmic fraction $f_{\rm H,c}$, $\beta_1$, the logarithmic slope of the relation, and $v_{\rm c,0}$ (in km/s), which represents the minimum circular velocity of haloes able to host HI. The best-fitting values and uncertainties on these parameters are constrained by a joint fit to all the currently available HI data (i.e. observations of DLAs, HI galaxies, and intensity mapping experiments). Forecasting of cosmological parameters $\{h, \Omega_m, n_s, \Omega_b, \sigma_8\}$ within such a framework \citep{padmanabhan2019} yields excellent results, indicating that CHORD may have the ability to simultaneously improve our knowledge of the astrophysics of HI and place stringent constraints on cosmology.

\begin{figure}
    \centering
    \includegraphics[width = 0.75\columnwidth]{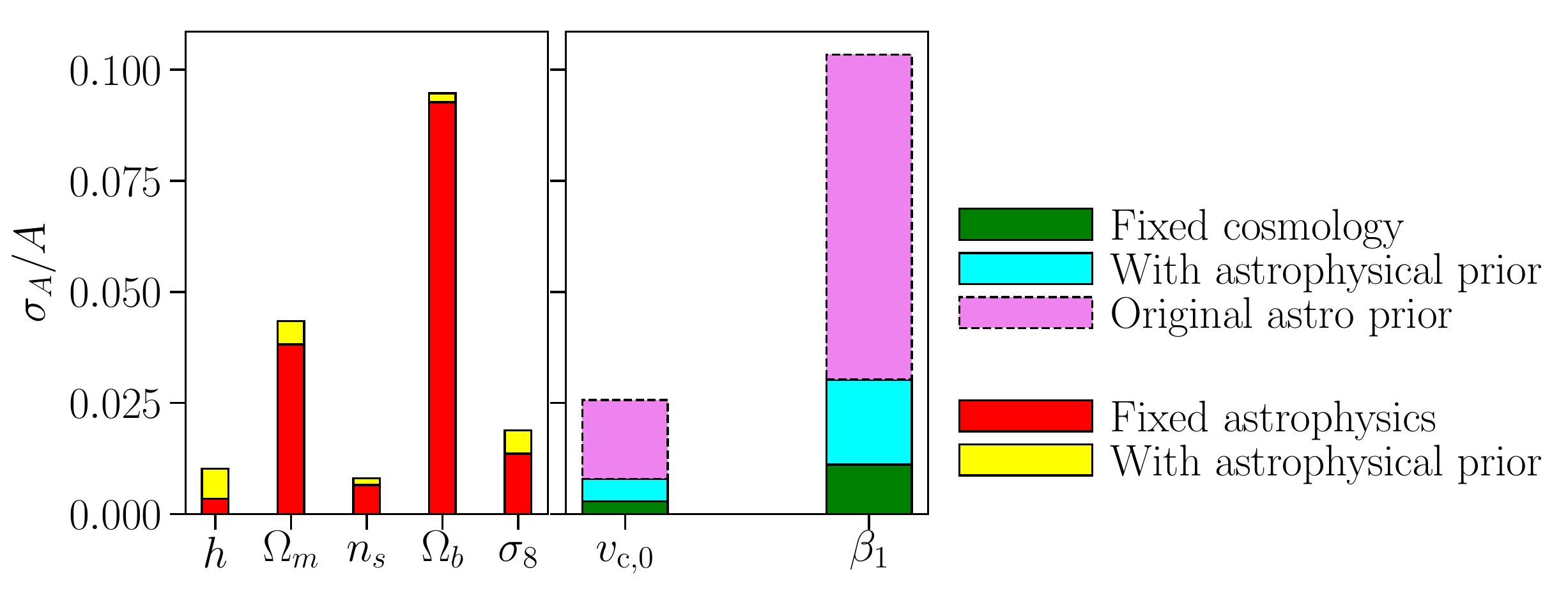} 
    \caption{Constraints on cosmological and astrophysical parameters using a CHORD-like survey over $z \sim 0-3$, following the approach developed in \citet{padmanabhan2019} based on the halo model framework for HI in \citet{padmanabhan2017a}. Noise parameters correspond to a fiducial CHORD configuration of 512 3-m dishes representing the cosmology core. Foregrounds are added following \citet{santos2005} with an optimal level of foreground subtraction appropriate for large array configurations \citep[$\epsilon_{\rm FG} = 10^{-6}$ in the notation of][]{bull2015}. {\textit{Left panel:}} Constraints on the cosmological parameters $\{h, \Omega_m, n_s, \Omega_b, \sigma_8\}$ assuming a flat $\Lambda$ CDM framework, with fixed astrophysics parameters (red) and marginalizing over astrophysics (yellow). {\textit{Right panel:}} Constraints on the cutoff, $v_{\rm c,0}$ and the logarithmic slope $\beta_1$ of the HI-halo mass relation in Eq. \ref{eq:HIhalomodel}, for fixed cosmology (green) and marginalizing over cosmological parameters but adding an astrophysical prior (cyan). The extent of the current astrophysical prior is plotted as the violet band in each case.}
    \label{fig:chordforecasts}
\end{figure}

\item Future 21 cm surveys e.g., with CHORD, should be able to detect the ``differential neutrino condensation effect" \citep{yu2017}, in which the neutrino richness of a region affects the halo mass function, in a way that is separable from baryonic effects. For a fiducial sum of neutrino masses of 0.05 eV, such surveys can also enable up to a 5$\sigma$ detection of the effect of tidal torque caused by neutrinos on dark matter haloes \citep{yu2019}.
The relative velocity between relic neutrinos and dark matter at low redshifts leads to a neutrino wake, downstream of the dark matter haloes. This causes a dipole distortion of galaxy-galaxy lensing, detectable from cross-correlations between different galaxy populations and 21 cm intensity mapping experiments \citep{okoli2017, zhu2014}. High precision 21 cm lensing surveys (e.g. with CHORD) may even be able to distinguish normal and inverted neutrino hierarchies, and detect right-handed Dirac neutrinos using this effect \citep{zhu2016}.

\item 
At larger scales and earlier redshifts, the cosmic density field has undergone less ``processing" by nonlinear clustering and astrophysical evolution than at smaller scales and later times. Thus, observations in the former regime (where future \tcm surveys will have unparalleled performance) can be more easily related to the physics of the early universe. Constraints on two of the key observational targets for learning about this era---signatures in multi-point density correlation functions that depend on the field content of the early universe (so-called ``primordial non-Gaussianity" \citealt{Alvarez:2014}), and sharp or oscillatory features in the primordial power spectrum that generically arise in many models of cosmic inflation~\citep{Chluba:2015bqa}---scale with the number of accessible linear Fourier modes (see Figure~\ref{fig:modes}). \tcm surveys, as the most cost-effective and scalable way of surveying large-scale structure, are ideally positioned to capitalize on this scaling and vastly improve our knowledge of the early universe, as verified in recent forecasts~\citep{Ansari:2018}.
\item 
\tcm fluctuations will be distorted by gravitational lensing, and these distortions can be used to reconstruct projected maps of the intervening large-scale structures that act as lenses, by adapting techniques that have been developed and applied to observations of the cosmic microwave background (e.g.~\citealt{Foreman:2018}). \tcm measurements, with their naturally fine redshift resolution, can be divided into a sequence of ``screens," with lensing of each screen probing a different line-of-sight projection (i.e.\ redshift weighting) of the lower-redshift lenses. These lensing measurements will be highly complementary to those from optical lensing surveys, as they will have complementary systematics and also provide screens at higher redshifts.
\end{itemize}


\section{Opportunities and Recommendations for Canada in the coming decade}
\label{sec:CHORD}

Capitalizing on the opportunities described in Section \ref{sec:Frontiers} will require continued and sustained investment in $21\,\textrm{cm}$ instrumentation, analysis, theory, and other complementary probes of our Universe:

\subsection{The Canadian Hydrogen Observatory and Radio transient Detector (CHORD)}

CHORD is a proposed next-generation radio observatory, building off the legacy and lessons learned from instruments such as CHIME. (See CHORD white paper E029 for details). The CHORD project consists of three phases. Phase I adds outrigger cylinders to CHIME, and Phase II similarly adds wideband outrigger dishes to CHIME. The long baselines provided by these outriggers enable a precise localization of radio transients such as Fast Radio Bursts. For the cosmological applications described in Section \ref{sec:Frontiers}, the most important part of CHORD will be its third phase. In Phase III, an array of 512 close-packed 6-m dishes will be constructed. Each dish will be fitted with wideband receivers covering the range $300\,\textrm{MHz}$ to $1.5\,\textrm{GHz}$.

\textbf{We recommend the funding of CHORD, or a CHORD-like telescope.} CHORD represents a sensible next-generation instrument for \tcm cosmology for several reasons. First, it possesses a large collecting area, providing it with sufficient sensitivity to perform the beyond-BAO measurements highlighted in Section \ref{sec:Frontiers}. Second, its broadband receivers enable a long lever arm in redshift, pushing low-$z$ \tcm cosmology to redshifts that other cosmological probes have difficulty reaching. Finally, CHORD builds on the lessons learned from current-generation experiments such as CHIME. The result is a much finer control of systematics. For example, CHORD's composite dishes will be much more precisely engineered than CHIME's cylinders. With a surface precision of a thousandth of a wavelength, the assumption that different elements of an interferometer have identical beams will be a much more appropriate one with CHORD. This minimizes the need to take out systematic effects in software, where slight mis-modelings can result in residuals that are much larger than the faint cosmological signals.

\subsection{Small-scale hardware tests}

The low-cost manufacture of CHORD's precise composite dishes is just one example of a technological advance that makes large, next-generation arrays like CHORD possible. In addition to dish manufacturing, CHORD also benefits from advances in ultra-wideband feed design, ultra-low noise amplifiers, and correlator technology. \textbf{We recommend that small-scale investments continue to be made in the development of next-generation radio hardware.} This will not only position the Canadian community for future generations of radio interferometers, but may also benefit high-$z$ \tcm cosmology. Despite being in a different redshift range, many shared technologies exist, and at high redshifts, relatively small instruments seeking to measure the angularly averaged \tcm signal can have an enormous impact. (See white paper E011).

\subsection{Complementary observations}

As evidenced by the early results from GBT and Parkes, cross-correlations are not only a powerful way to evade foreground and instrumental systematics, but also yield interesting scientific constraints in their own right. Leveraging existing Canadian expertise in cross correlating with galaxy surveys, a promising future opportunity is to cross correlate with data from the Dark Energy Spectroscopic Instrument (DESI). DESI will obtain spectroscopic measurements of galaxies at $z \lesssim 1$ and quasars at $z \sim 3.5$, providing a crucial connection between \tcm intensity mapping experiments and larger international galaxy surveys.

\textbf{Additionally, we recommend that Canada pursue opportunities to survey our Universe with non-\tcm lines.} Examples of such lines include CO rotational lines and the [CII] fine structure line. Cross correlations with different tracers of large scale structure will again not only mitigate systematics, but will also shed light on detailed galaxy formation properties (see, e.g., \citealt{Breysse2019AGN} for an example of how AGN feedback can be probed CO-quasar cross correlations). Please see white paper E069 for more detailed recommendations.

\subsection{Participation in international next-generation efforts.} Canadian efforts such as CHIME and CHORD will produce significant science results in the first half of the next decade. Crucially, however, they also serve as important stepping stones for large-scale international efforts that will become important in the second half of the next decade. An example of this would be the Packed Ultra-wideband Mapping Array (PUMA; \citealt{PUMA}), which is similar in design and science goals to CHORD. Canadian scientists from CHORD will therefore be well-poised to assume leadership positions within PUMA. The same will be true for the SKA. \textbf{We recommend that Canadian scientists be funded to continue participating in these long-term efforts, whether through analysis, theory, or hardware contributions.}

\subsection{Theory and Technical Support.} As evidenced by Section \ref{ref:Theory}, Canadian astronomers have provided many of the key theoretical and analysis advances in the field of \tcm cosmology. These avenues of research will become even more important going forward, now that data is available and increasing emphasis is placed on analysis and interpretation. \textbf{We therefore recommend continued support for institutions such as CITA.}

In addition, the development and maintenance of infrastructure (including software) for projects such as CHIME and CHORD often requires a dedicated staff comprised of non-tenure-track/non-academic scientists and engineers. \textbf{We recommend that support for such staff continue, as it is vital not only for the smooth running of large experimental efforts, but also in the development of Canadian technical leadership.}




\begin{lrptextbox}[How does the proposed initiative result in fundamental or transformational advances in our understanding of the Universe?]
The \tcm line is a transformational new probe of cosmology, opening up a new redshift regime to maps of large-scale structure. While individual astrophysical objects have been detected and studied at the redshifts covered by low-$z$ \tcm experiments, the proposed initiatives will represent the first efforts at systematically probing large-scale structure up to $z \sim 6$. As a first step, BAO measurements with telescopes such as CHIME and CHORD will provide insights into the nature of dark energy using observations that have independent systematics to those of galaxy surveys. Should this produce interesting phenomenology beyond a ``vanilla" cosmological constant, this would be transformational and particularly timely, given current tensions in the Hubble parameter. Beyond BAOs, the \tcm line will also deliver precision cosmological constraints beyond the standard model (including but not limited to the detection of primordial non-Gaussianities or measurements of the sum of the neutrino masses). Such constraints will be aided by the ability to go beyond the $z \sim 1$ regime that is the reach of many traditional probes of cosmology---at higher redshifts, more modes are in the linear regime, making modelling simpler. Finally, \tcm measurements will shed light on galaxy formation, as evidenced by the fact that puzzles are already emerging in the interpretation of existing cross-correlations measurements.
\end{lrptextbox}

\begin{lrptextbox}[What are the main scientific risks and how will they be mitigated?]
Ideally, the \tcm line will be used both to make HI autocorrelation measurements as well as cross correlation measurements with other line intensity maps or galaxy surveys. To date, however, positive detections have been made only of the latter, and only upper limits exist for the former. This is due to the technical difficulties of measuring a tiny cosmological from underneath foreground contaminants that are many orders of magnitude brighter. This contamination is not only a problem in its own right, but also places stringent requirements on the control of instrumental systematics. The problems of foregrounds and systematics therefore pose a risk to the program we have outlined. However, there are reasons for optimism. First, CHIME and HIRAX will serve as pathfinders that will retire the risk for larger scale projects like CHORD. Indeed, with large amounts of data now available, foregrounds and systematics can now be tackled head-on in an empirical fashion. Moreover, one advantage of building an interferometer is that data can be taken with a partial array, and necessary hardware adjustments can be made if initial data for a telescope do not seem to meet specifications.
\end{lrptextbox}

\begin{lrptextbox}[Is there the expectation of and capacity for Canadian scientific, technical or strategic leadership?] 
There is a strong expectation of Canadian leadership, given that CHIME is located in Canada and the Principal Investigators of both CHIME and HIRAX are based at Canadian institutions. CHORD, similarly, will be a Canadian telescope that is located in Canada (aside from possibly a few outriggers stations), and will be led almost entirely by Canadians. No other low-redshift \tcm experiments of this scale exist anywhere in the world, making Canada the undisputed leader in \tcm cosmology at $z < 6$.
\end{lrptextbox}

\begin{lrptextbox}[Is there support from, involvement from, and coordination within the relevant Canadian community and more broadly?] 
The Canadian \tcm community is not limited to low-redshift efforts, and there is considerable overlap (and sharing of hardware and analysis techniques) with the high-redshift \tcm community. Additionally, the Dominion Radio Astronomy Observatory (DRAO) provides site support for CHIME and CHORD. Infrastructure support for the ``big data" nature of modern radio astronomy is also provided by efforts such as CIRADA.
\end{lrptextbox}

\begin{lrptextbox}[Will this program position Canadian astronomy for future opportunities and returns in 2020-2030 or beyond 2030?] 
This program enables Canadian astronomy to play a key role in enabling (and mitigating risk) for international efforts later in the 2020-2030 decade and beyond. The PUMA array \citep{PUMA} is being proposed in the US, and is conceptually reasonably similar to a larger version of CHORD. Lessons learned from CHORD will therefore have a substantial influence on PUMA, and the eventual science case for PUMA will certainly be affected by the portfolio of results coming out of CHORD. Canadian astronomers will therefore be key members in the PUMA collaboration (indeed, some already are), using their experience to lead international efforts to groundbreaking science results. Canadians will be in a similar situation for the SKA, and will likely have a larger influence than the US community given that they are not formally involved with the project.
\end{lrptextbox}

\begin{lrptextbox}[In what ways is the cost-benefit ratio, including existing investments and future operating costs, favourable?] 
\tcm intensity mapping is an extremely cost-efficient way to map large-scale structure, in principle allowing huge volumes to be surveyed at a fraction of the cost of traditional galaxy surveys. In addition, Canada benefits from existing investments in instruments such as CHIME, as well as site and operations support from DRAO.
\end{lrptextbox}

\begin{lrptextbox}[What are the main programmatic risks
and how will they be mitigated?] 
While CHIME is fully funded, HIRAX is only partially funded, and CHORD is still being proposed. There is thus some risk that investments in next-generation telescopes will slow down. Many non-\tcm line intensity mapping experiments that might serve as cross-correlation targets are also in the state of being only partially funded, resulting in some risk regarding the availability of other lines to cross correlate with.
\end{lrptextbox}

\begin{lrptextbox}[Does the proposed initiative offer specific tangible benefits to Canadians, including but not limited to interdisciplinary research, industry opportunities, HQP training,
EDI,
outreach or education?] 
As a relatively young field that has tended towards the model of building experiments rather than applying for time on general observatories, \tcm cosmology is a field where hardware opportunities are abundant for HQP training. The large datasets generated by typical \tcm experiments also provide abundant opportunities for collaborations with computer scientists and those researching ``big data" methods.
\end{lrptextbox}
\setlength{\bibsep}{0.0pt}
\bibliography{example} 

\end{document}